\documentclass[a4paper,fleqn,usenatbib]{mnras}

\usepackage{mathptmx}
\usepackage{graphicx, color}
\usepackage{ulem}

\usepackage[T1]{fontenc}
\usepackage{ae,aecompl}

\usepackage{graphicx}	
\usepackage{amsmath}	
\usepackage{amssymb}	

\title[Dust reverberation mapping of type 2 AGN]
{
Dust Reverberation Mapping  of Type 2 AGN NGC~2110 Realized with X-ray and 3--5~$\mu$m IR monitoring
}

\author[H.~Noda et al.]{
Hirofumi Noda$^{1,2}$\thanks{E-mail: noda@ess.sci.osaka-u.ac.jp}, 
Taiki Kawamuro$^{3}$, 
Mitsuru Kokubo$^{4}$, 
Takeo Minezaki$^{5}$
\\
$^{1}$ Department of Earth and Space Science, Graduate School of Science, Osaka University, 1-1 Machikaneyama, Toyonaka, Osaka 560-0043, Japan\\
$^{2}$ Project Research Center for Fundamental Sciences, Osaka University, 1-1 Machikaneyama, Toyonaka, Osaka 560-0043, Japan\\
$^{3}$ National Astronomical Observatory of Japan, Osawa, Mitaka, Tokyo 181-8588, Japan\\
$^{4}$ Astronomical Institute, Tohoku University, 6-3 Aramakiazaaoba, Aoba-ku, Sendai, Miyagi 980-8578, Japan\\
$^{5}$ Institute of Astronomy, School of Science, the University of Tokyo, 2-21-1 Osawa, Mitaka, Tokyo 181-0015, Japan
}

\date{Accepted 2020 May 13. Received 2020 April 19; in original form 2019 September 29}

\pubyear{2015}

\begin{document}
\label{firstpage}
\pagerange{\pageref{firstpage}--\pageref{lastpage}}
\maketitle

\begin{abstract}
The dust reverberation mapping is one of powerful methods to investigate the structure 
of  the dusty tori in  AGNs, and it has been performed on more than a hundred type 1 AGNs. 
However, no clear results have been reported on type 2 AGNs because their strong 
optical--UV extinction completely hides their accretion disc emission. 
Here we focus on an X-ray-bright type 2 AGN, NGC~2110, and utilize 
2--20~keV X-ray variation monitored by \textit{MAXI} to trace disc emission, 
instead of optical--UV variation. 
Comparing it with light curves in the \textit{WISE} infrared (IR) W1 band ($\lambda=3.4~\mu$m) 
and W2 band ($\lambda=4.6~\mu$m)  with cross-correlation analyses, 
we found candidates of  the dust reverberation time lag at $\sim 60$~days, $\sim 130$~days, 
and $\sim 1250$~days between the X-ray flux variation and those of the IR bands. 
By examining the best-fitting X-ray and IR light curves with the derived time lags, 
we found that the time lag of $\sim 130$~days is most favoured. 
With this time lag, the relation between the time lag and luminosity of NGC~2110 
is consistent with those in type 1 AGNs, suggesting that the dust reverberation in NGC~2110 
mainly originates in hot dust in the torus innermost region, the same as in type 1 AGNs. 
As demonstrated by the present study, X-ray and IR simultaneous monitoring can 
be a promising tool to perform the dust reverberation mapping on type 2 AGNs. 
\end{abstract}



\section{Introduction}


 The dusty torus is an important structure in the active galactic nucleus (AGN) 
and is believed to surround an accretion disc, which lies in the immediate vicinity 
of  the central supermassive black hole (SMBH), and a broad line region (BLR). 
According to the classical AGN unified model (e.g., \citealt{1993ARA&A..31..473A}; \citealt{1995PASP..107..803U}), 
all AGNs have the same structure, but appear differently  depending on  the viewing angle 
(the  angle of the line of sight against  the rotation axis of  the torus), which determines 
the differences between the observational classification of the two types of AGNs, types 1 and 2.
The AGNs that show optical broad emission lines with a velocity width of $\gtrsim 2000$~km~s$^{-1}$
 are  classified as type 1,  whereas those  do not are  as type 2. 
 The unified model explains that their viewing angles are different so
 that the accretion disc and BLR are  obscured in the  latter but not in the former.
 AGN dust tori have been usually studied through analyses of their infrared (IR) 
 spectral energy distributions (SEDs), along with theories developed to explain them. 
According to the most recent models of the dust torus,
the IR continuum emission has multiple emission components, 
as opposed to a single component, which was once assumed according to the conventional models.
The near infrared (NIR) continuum emission, which presumably originates from 
hot dust residing in the torus innermost region, is now considered to be a 
separate emission component, and hence, the geometry of the torus innermost region
and its relation to neighboring AGN structures such as the accretion disc and the BLR 
are under much debate 
(e.g., \citealt{2009ApJ...705..298M}; \citealt{2017ApJ...835..257L};
\citealt{2017ApJ...838L..20H}; \citealt{2018MNRAS.474.1970B}).

The dust reverberation mapping  is one of 
the most powerful tools to investigate the compact structure of the torus innermost region of type 1 AGNs
and  has been applied to various of them since 1970s 
(e.g., \citealt{1974MNRAS.169..357P}; \citealt{1980Natur.284..410L}; \citealt{1989ApJ...337..236C}).
When the accretion-disc continuum flux in the optical to ultraviolet (UV) bands varies,
the flux of the  innermost region of the torus in NIR band
responds with dust reverberation time lags $\tau_{\rm dust}$, 
 with which  the distance, i.e. radius, from  the SMBH to the torus innermost-region
can be estimated as $R_{\rm dust}=c\tau_{\rm dust}$.
 Since $R_{\rm dust}$ is considered to be close to  the radius
 where it is at the dust sublimation temperature of
$1700$--$2000$~K (for graphite grains:
\citealt{1977AdPhy..26..129H}; \citealt{1977ARA&A..15..267S}; \citealt{2018MNRAS.474.1970B}),
$R_{\rm dust}$ is expected to  be proportional to the square-root of the disc luminosity,
 $R_{\rm dust}\propto L_{\rm disc}^{0.5}$ (\citealt{1987ApJ...320..537B}).
Large systematic dust-reverberation surveys for type 1 AGNs
indeed demonstrated that the dust-reverberation radius approximately
follows the radius-luminosity relation of $R_{\rm dust}\propto L_{\rm disc}^{0.5}$
 for a wide luminosity range
(\citealt{2006ApJ...639...46S}; \citealt{2014ApJ...788..159K};
 \citealt{2019ApJ...886...33L}; \citealt{2019ApJ...886..150M}).
Recent advancements of NIR interferometers  directly revealed the extent of 
the innermost dust torus  and confirmed  its luminosity dependency
 (e.g., \citealt{2011A&A...527A.121K}; \citealt{2019arXiv191000593G}).

However, no clear results of the dust reverberation mappings have been 
reported  about type-2 sources, which dominate  the population of  AGNs,
 mainly because their optical/UV emission is heavily or even totally  absorbed, 
 rendering monitoring their disc-flux variation difficult or even impossible at the optical or UV wavelengths.
 A potential way to circumvent this problem with type 2 AGNs is to observe X-rays from a corona  lying at an inner part of 
 the accretion disc near  the central SMBH as a tracer of disc-flux variation, instead of optical or UV  light.   
Long-term X-ray variation of AGNs is known to show a good correlation with optical/UV variation with 
an  time lag of hours to days for optical/UV lights behind X-rays, 
as revealed by X-ray and optical/UV monitoring on various type-1 AGNs 
(e.g., \citealt{2014ApJ...788...48S}; \citealt{2016ApJ...828...78N}). 
The correlation between the optical/UV and  X-rays 
can be  more or less explained as disc reverberation by X-ray irradiation from a corona  to the disc 
(e.g., \citealt{1991ApJ...371..541K}; \citealt{2015ApJ...806..129E}; \citealt{2016ApJ...828...78N}). 
 Given that the disc-reverberation time lag between 
X-rays and optical/UV  is negligible compared  with $\tau_{\rm dust}$,  precise 
estimate of $\tau_{\rm dust}$  with X-ray and NIR monitoring should be feasible. 
Therefore, we expect that simultaneous X-ray and NIR monitoring observations 
enable us to realize the dust reverberation mapping of type 2 AGNs for the first time. 

In the present paper, we analyze 2--20~keV monitoring data  with Monitor of All-sky X-ray Image 
(\textit{MAXI}; \citealt{2009PASJ...61..999M}) and the W1 ($3.4~\mu$m) and W2 ($4.6~\mu$m) band photometry data  with 
Wide-field Infrared Survey Explorer (\textit{WISE}; \citealt{2010AJ....140.1868W}) on  the type 2 AGN NGC~2110 (R.A=5$^{\rm h}$~52$^{\rm m}$~11.4$^{\rm s}$, Dec=$-7^{\rm d}$~27$^{\rm m}$~22$^{\rm s}$), which is well known and 
X-ray bright \citep{2018ApJS..238...32K}. 
NGC~2110 is  a nearby Compton-thin type-2 Seyfert galaxy 
 with a redshift of 0.0078  at a distance  of 31.2~Mpc \citep{1999MNRAS.304...35S}. 
Two  independent measurements of the mass of the SMBH at its  centre were reported;  
\cite{2002ApJ...579..530W} showed $M_{\rm BH} = 2.0\times10^8~M_{\odot}$, using the relation between 
$M_{\rm BH}$ and  stellar velocity dispersion,  whereas \cite{2015ApJ...802...98M} reported 
$M_{\rm BH} = 2.5\times10^7~M_{\odot}$, using the velocity width of the narrow Fe-K$\alpha$ line. 
We adopt  cosmological parameters of $H_0 = 73$~km~s$^{-1}$~Mpc$^{-1}$, 
$\Omega_{\Lambda} = 0.73$, and $\Omega_{\rm m} = 0.27$ throughout this paper. 
Errors shown in text and figures refer to those in $1\sigma$ confidence, unless otherwise stated.

\section{Observations and Data Reduction}

\subsection{X-ray data}
\textit{MAXI} onboard the International Space Station (ISS) has been operated from 2009 August 15, 
  continuously monitoring all sky  with the Gas Slit Cameras (GSCs; \citealt{2011PASJ...63S.623M}) 
  in 2--30~keV and the Solid-state Slit Cameras (SSC; \citealt{2011PASJ...63..397T}) in 0.5--12~keV.  
  Since soft X-rays of NGC~2110 are  heavily absorbed by the torus \citep{2016ApJS..225...14K}, 
we  use hard X-ray data above 2~keV obtained by the GSCs only 
in  this work. 
The GSCs cover two 84$^{\circ}$-separated regions with fields of view of $3^{\circ}.0 \times 160^{\circ}$ 
and observe individual  areas of the sky every 92~minutes  synchronized with the rotation of the ISS. 
The GSCs can observe 85\% and 95\% of  all sky per one ISS rotation and one day respectively.  
Thus NGC~2110  has been monitored almost everyday. 

\begin{figure*}
	\includegraphics[width=140mm]{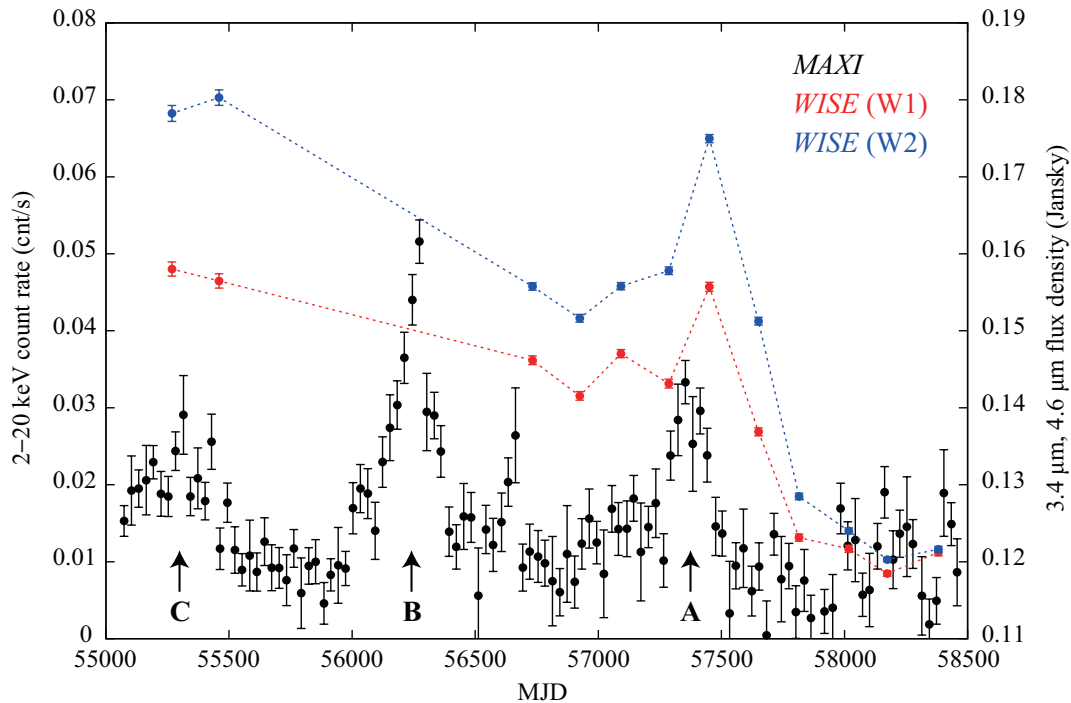}
    \caption{ Light curves in (black) 2--20~keV  with \textit{MAXI}/GSC  and  the (red) W1  and (blue) W2 bands 
     with \textit{WISE}. The dotted lines  simply connect the W1 and W2 individual data points. The X-ray flares at 57250--57500~MJD, 56000--56500~MJD and 55000--55500~MJD are named  peaks A, B and C, respectively. }
    \label{fig:fig1}
\end{figure*}

\begin{figure}
	\includegraphics[width=85mm]{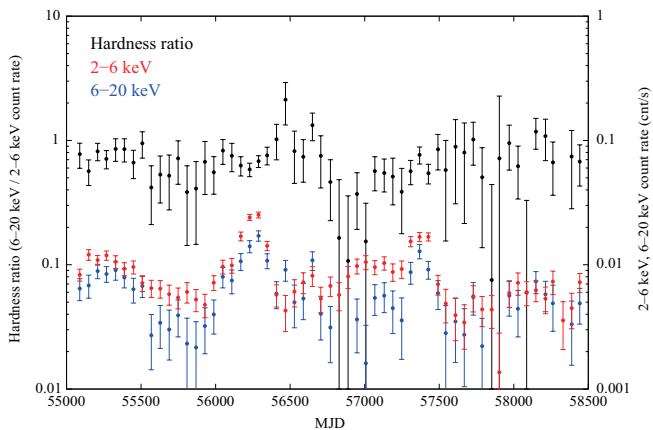}
    \caption{Light curves  in the (red) 2--6~keV  and (blue) 6--20~keV bands, and a curve of (black) their ratio (hardness ratio).}
    \label{fig:fig2}
\end{figure}

\begin{figure*}
	\includegraphics[width=160mm]{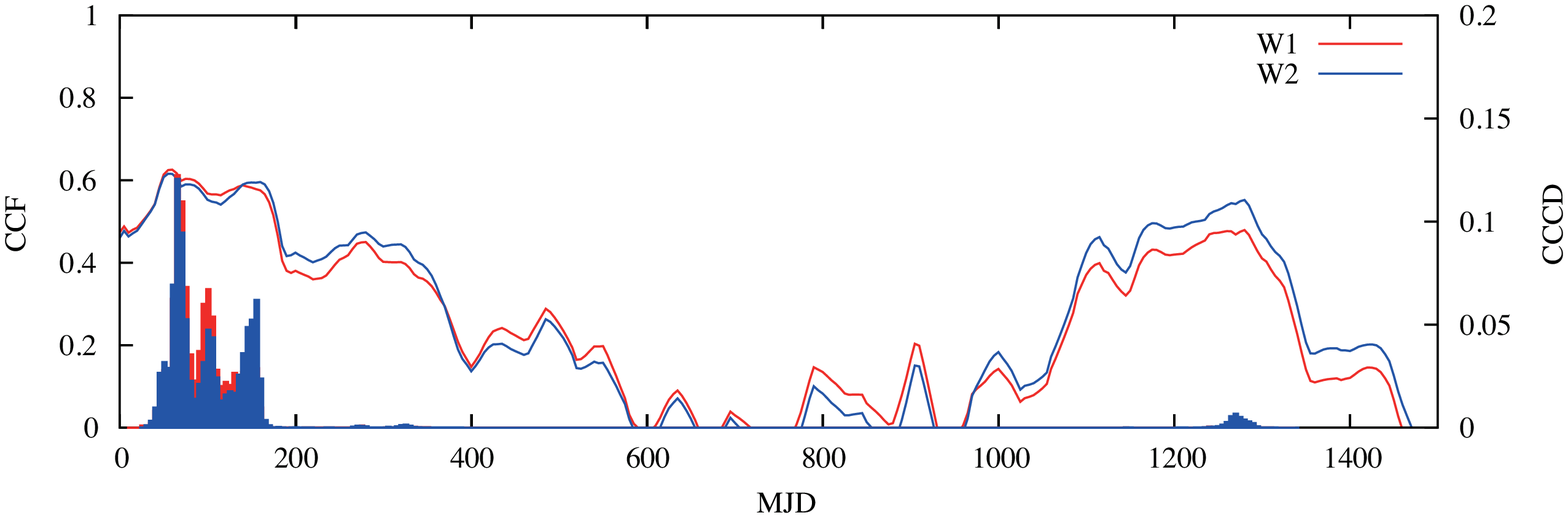}
    \caption{ ICCFs between the 2--20~keV and the (red) W1 and (blue) W2 bands  data, along with the histogram of their CCCD. }
    \label{fig:fig3}
\end{figure*}

Using the on-demand pipeline provided by the  \textit{MAXI} team (http://maxi.riken.jp/mxondem/), 
we reduced the GSC data of NGC~2110 obtained from 55000 to 58500~MJD and extracted a 30-days binned light curve 
in the 2--20~keV band. We also extracted 60-days binned light curves in the 2--6~keV and 6--20~keV bands 
and made a curve of a hardness ratio  of the  6--20~keV to 2--6~keV fluxes, using the default on-demand  interface. 

\subsection{3--5~$\mu$m IR data}
 \textit{WISE} is an astronomical infrared satellite, launched on 2009 December 14, and performed 
 all-sky survey until 2011 at the 3.4, 4.6, 12, and 22~$\mu$m bands  named  W1, W2, W3, and W4, 
respectively. The programme \textit{Near-Earth Objects WISE} (\textit{NEOWISE}) 
  \citep{2014ApJ...792...30M} started in 2013 after the initial and primary programme with use of cryogen 
  and has been running since then, after a long period of interruption from 2011 February to 2013.

In  this work, we  used the \textit{AllWISE} multi-epoch photometry table and \textit{NEOWISE} single exposure 
(L1b) source table  fetched from the NASA/IPAC Infrared Science Archive (https://irsa.ipac.caltech.edu/Missions/wise.html). 
We  obtained the photometry results in an area within a $10"$ radius 
of (RA, Dec) = ($88.04742$, $-7.45621$) in the J2000 coordinates, 
and took the values of the W1 magnitude \texttt{w1mpro\_ep},  
W1 magnitude error \texttt{w1sigmpro\_ep},  W2 magnitude \texttt{w2mpro\_ep},  
W2 magnitude error \texttt{w2sigmpro\_ep}, and  MJD \texttt{mjd}. 
The \textit{WISE} magnitude values  were converted to the flux density $F_{\nu}$  
according to the equation  $F_{\nu} = F_{\nu, 0} \times 10^{(-m/2.5)}$, 
where $F_{\nu, 0} =  309.540$~Jy and 171.787~Jy for the W1 and W2 bands, 
respectively, and $m$ is the \textit{WISE}  magnitude.  
After the conversions,  the averages of the flux densities per visit were 
calculated and accepted as the W1 and W2 fluxes in each visit.

\section{Data Analysis}

\subsection{X-ray and 3--5~$\mu$m flux correlation with dust-reverberation time lag}

Figure~\ref{fig:fig1} shows long-term light curves in the 2--20~keV band  with \textit{MAXI} 
and in the W1 and W2 bands  with \textit{WISE}. 
 A significant variation in the 2--20~keV  band was  detected with an unprecedentedly 
 high-cadence observations with \textit{MAXI}. 
 The most remarkable was three  distinctive flares at 57250--57500~MJD,  
 56000--56500~MJD, and 55000--55500~MJD from  later to  earlier; 
 hereafter we  refer to them as the (X-ray) peaks A, B, and C, respectively, as marked in Fig.~\ref{fig:fig1}.
The 2--20~keV flux  varied by a factor of
$\sim 3$ and $\sim 5$ in  peaks A and B, respectively,  each lasting hundreds of days.

This kind of long-term X-ray flux  variation is considered to show a good correlation 
with  that of disc emission in optical/UV band (e.g., \citealt{2014ApJ...788...48S}; \citealt{2016ApJ...828...78N}; 
\citealt{2017ApJ...840...41E}), 
 whereas fast X-ray variation on a timescale of hours--days is known to be uncorrelated to optical/UV variation 
(e.g., \citealt{2017MNRAS.470.3591G}; \citealt{2017ApJ...840...41E})
\footnote{
\cite{2017ApJ...840...41E} suggested that X-ray flux variation on a timescale of months 
is correlated with optical/UV flux variation because of reprocess by  the warm disc emitting 
soft X-rays, which is observed as a soft X-ray excess component (e.g., \citealt{2011A&A...534A..39M}; \citealt{2011PASJ...63S.925N}; \citeyear{2013PASJ...65....4N}; \citealt{2018A&A...611A..59P}; \citealt{2018MNRAS.480.3898N}). 
As an alternative suggestion, \cite{2016ApJ...828...78N} discussed that an additional rapidly-varying X-ray component (\citealt{2011PASJ...63..449N}; \citeyear{2013ApJ...771..100N}; \citeyear{2014ApJ...794....2N}; \citealt{2016PASJ...68S..28M})
appears or disappears, affecting the X-ray--optical/UV correlation. }.
The bin size of the \textit{MAXI} light curve of $\sim 1$~month in Fig.~\ref{fig:fig1} 
is long enough to smear out fast X-ray variability to represent the long-term X-ray flux variation,
and hence, the X-ray light curve can be used as a tracer of the disc flux variation of NGC~2110. 

 Since NGC~2110 is a type 2 AGN, its X-ray flux could be highly variable when a column density of the obscurers changes. 
In order to examine whether the variability is intrinsic or originates from the column density change, 
we made light curves in the 2--6~keV and 6--20~keV band, 
and produced a curve of the hardness ratio between them as shown in Fig.~\ref{fig:fig2}. 
 Then we performed a fit to the curve of the hardness 
ratio with a constant  function to examine whether the hardness had variability or not and found 
 an acceptable fit of $\chi^2$/d.o.f.=63.1/57. 
 This shows that the variation of the hardness ratio  was not significant, and hence the X-ray 
 variations were not mainly because of a potential variation of the absorption column density. 

Figure~\ref{fig:fig1} also shows the W1- and W2-band light curves. 
 The period of peak A  was  covered by the \textit{NEOWISE} phase 
with intervals of $\sim 180$~days, whereas neither of them covered 
the period of 55500--56500~MJD around peak B.  
A flare was detected at around 57500~MJD, roughly coinciding with X-ray peak A. 
The amplitude of the IR flare was $\sim 10$\% in both the W1- and W2- bands. 
In the following  analyses, we  examine the significances of the IR flare responses
to  X-ray peaks A  and B.   
 Peak C is not considered in this analysis, because
the total length of the monitoring period is so limited that 
the correlations between the delayed X-ray light curves with 
delays of $\gtrsim 2000$~days and the IR light curves are difficult to be examined, 
where the number of the available data points would be too small
for the correlation analysis to be statistically meaningful.

\begin{figure*}
	\includegraphics[width=140mm]{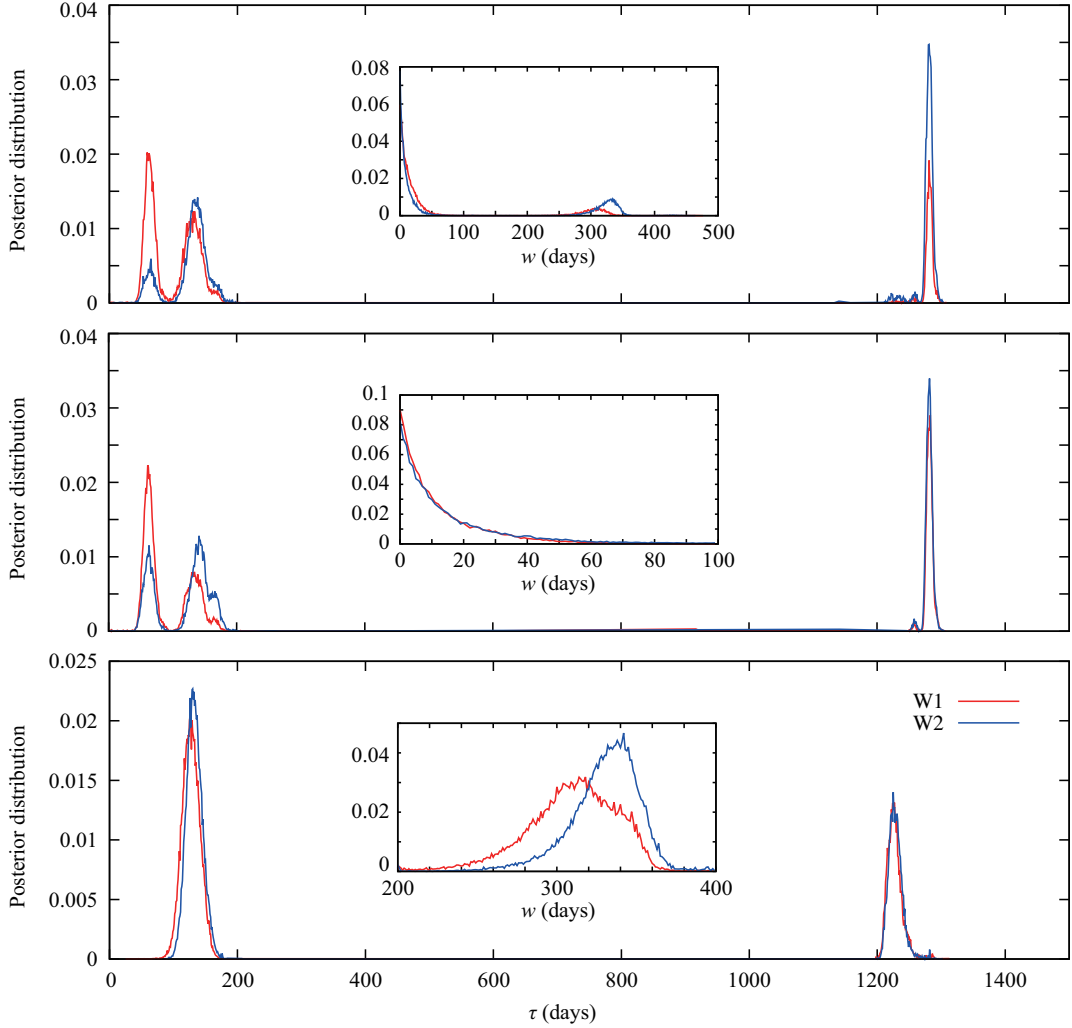}
    \caption{(top) Posterior distributions of $\tau$ in the W1 (red) and W2 (blue) bands behind the 2--20~keV emission obtained  with the JAVELIN algorithm without limiting $w$.  Inset in each panel shows  posterior distributions of $w$ derived  with the same JAVELIN run. (middle) Same as top panel, but by restricting $w$ within 0--100~days. (bottom) Same as top panel, but by restricting $w$ within 200--400~days. }
    \label{fig:fig4}
\end{figure*}

\begin{figure*}
	\includegraphics[width=165mm]{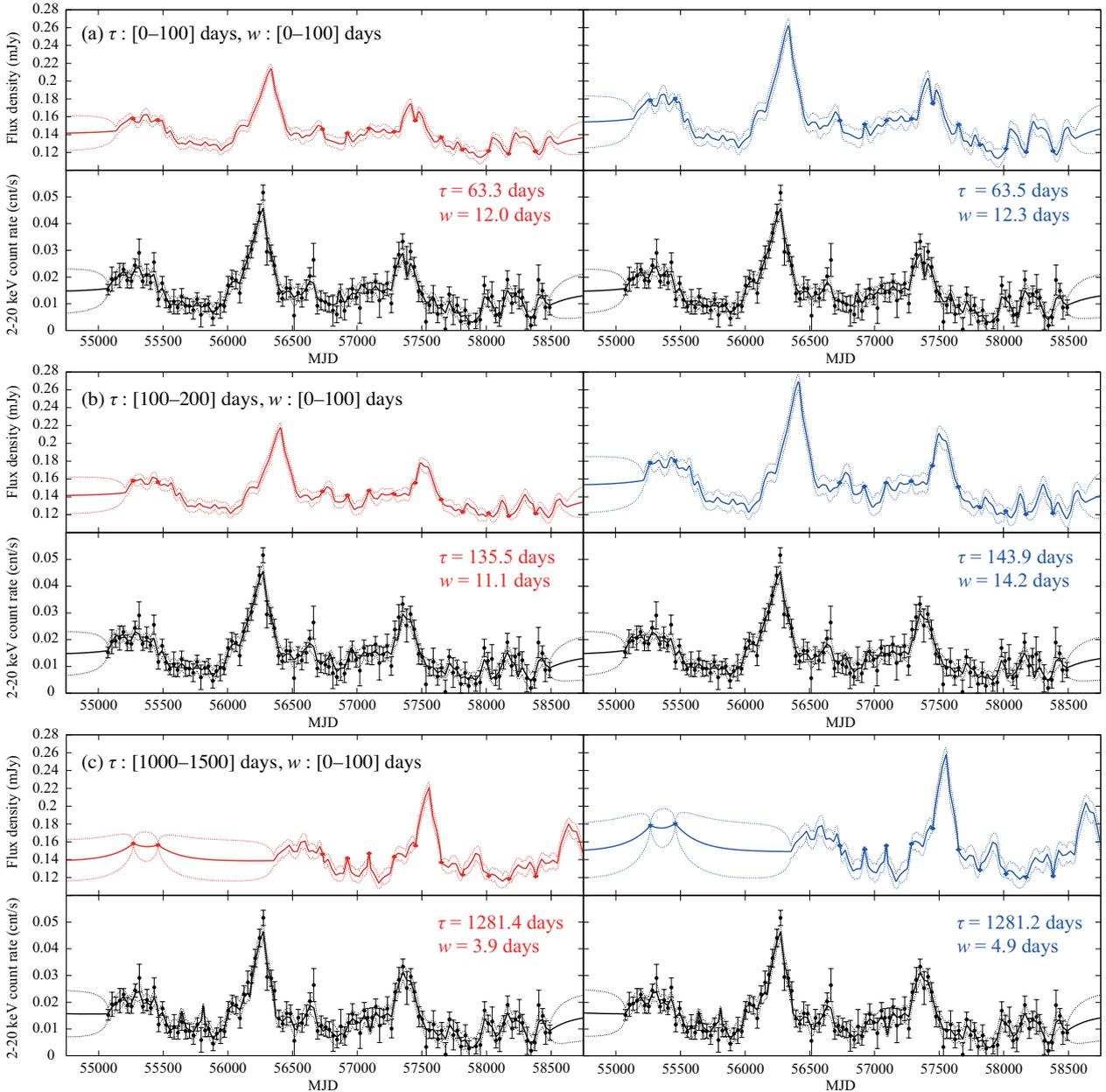}
    \caption{Best-fitting interpolated light curves (solid line) and their errors (dotted line) corresponding to the peaks in the top panel of Fig.~\ref{fig:fig4}, derived with the JAVELIN runs restricting $w$ within 0--100~days and $\tau$ within (a) 0--100~days, (b) 100--200~days, (c) 1000--1500~days.  In each set of 2$\times$2 panels, top two panels show the best-fitting light curves in the W1 (left panel, in red) and W2 (right panel, in blue) bands,  whereas the two bottom panels show those in 2--20~keV simulated with the (left panel) W1 and (right) W2 bands.  The best-fitting values of $\tau$ and $w$ are shown in each panel in the same  colours as the  light curves. }
    \label{fig:fig5}
\end{figure*}

\begin{figure*}
	\includegraphics[width=165mm]{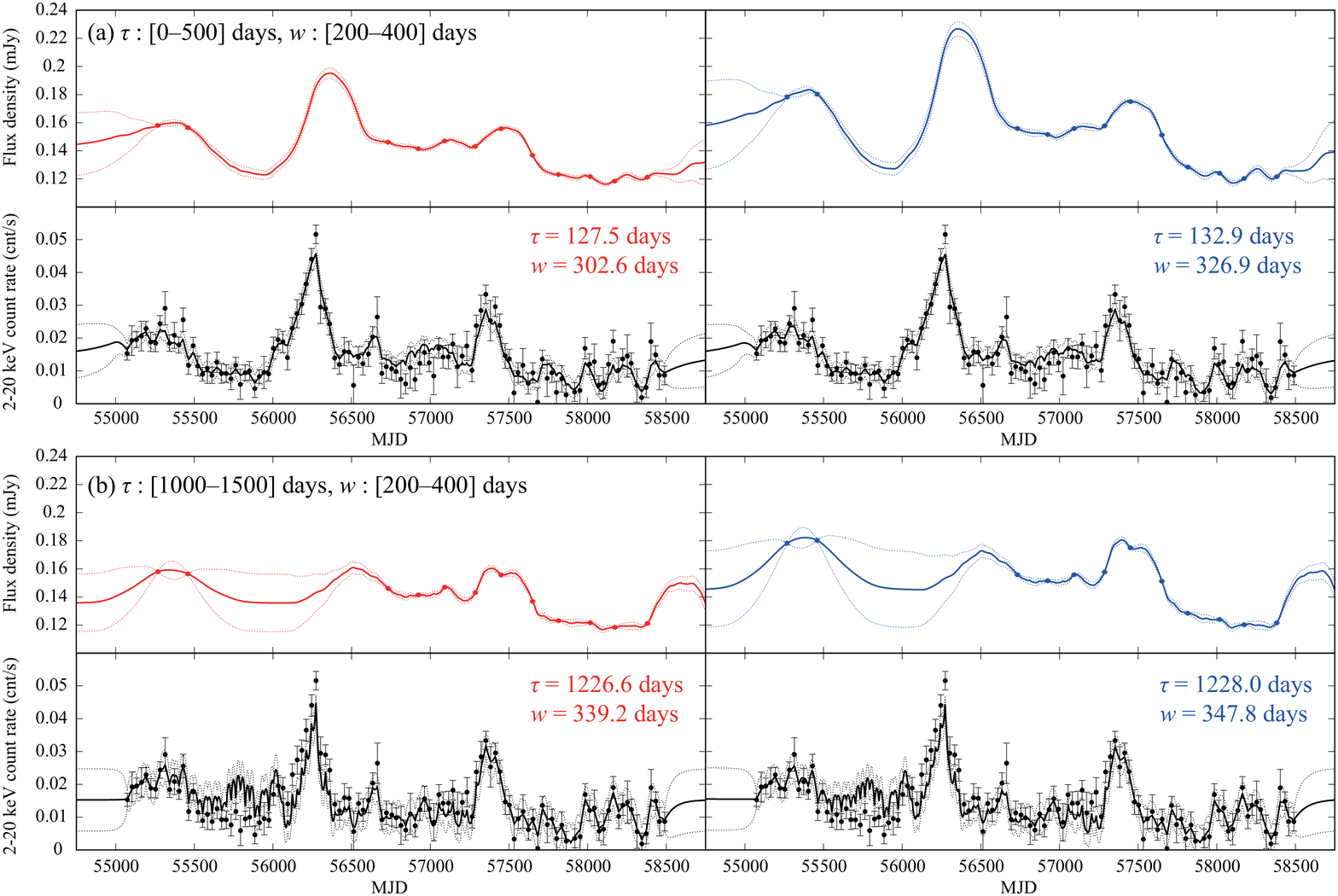}
    \caption{ Same as Figure~\ref{fig:fig5} but corresponding to the peaks in the bottom panel of Fig~\ref{fig:fig4} and for (a)  $\tau$ within 0--500~days and $w$ within 200--400~days, and (b)  $\tau$ within 1000--1500~days and $w$ within 200--400~days.
 }
    \label{fig:fig6}
\end{figure*}

In order to study the correlations between X-ray and the NIR flux variations, 
we conducted  analysis of the interpolation cross correlation function  \citep[ICCF;][]{1998PASP..110..660P} with 
a time lag $(\tau)$ range of 0--1500~days. 
In the ICCF analysis, the interpolation was applied to only the X-ray light curve with the time step of 5~days, 
because the X-ray sampling  was dense enough for the interpolation  whereas the IR samplings were sparse. 
In order to estimate the error of the time lags, the flux randomization method was 
employed with 30000 realizations in the Monte Carlo simulations\footnote{We used the code \texttt{pyCCF} in \url{http://ascl.net/code/v/1868}}.

Figure~\ref{fig:fig3} shows ICCF curves and cross correlation centroid distributions (CCCDs) 
derived  with the ICCF analysis. 
 Significant time lags $\tau$ in the W1 band ($\tau_{\rm W1}$) and W2 band ($\tau_{\rm W2}$) 
were  obtained to be $\sim 60$--150~days for both at $5\%$ significance level 
(correlation coefficients are $\sim 0.60$ with $n = 12$). 
This corresponds to the IR response to  X-ray peak A. 
The correlation coefficients were $\sim 0.47$ ($n = 12$) at $\tau_{\rm W1} = 0$ and
 $\sim 0.46$ ($n = 12$) at $\tau_{\rm W2} = 0$~days, 
 implying that the correlation with no time lag was not significant at 5\% significance level.
Another CCF peak  was found at $\tau\sim 1250$~days for both,
which corresponds to the IR response to  X-ray peak B.
However,  $\tau_{\rm W1}$ $\sim 1250$~days was not significant
even at $10\%$ significance level, 
whereas  $\tau_{\rm W2}$ $\sim 1250$~days was significant at
the same significance level,
from their correlation coefficients.

In order to further examine the time lags obtained  with the ICCF analysis, 
we employed the \texttt{JAVELIN} algorithm (\citealt{2011ApJ...735...80Z}; \citealt{2013ApJ...765..106Z}). 
The \texttt{JAVELIN} technique assumes a primary light curve  
to follow the dumped random walk (DRW) process (e.g., \citealt{2009ApJ...698..895K}) 
and convolves it with a top-hat transfer function (TF) to make
a responding light curve, searching with the Markov chain Monte Carlo (MCMC) method 
for the most likely values for the following five parameter: the amplitude and timescale of the DRW process, 
 height and width of the top-hat TF, and  response time delay. 
In our analysis, we assigned the 2--20 keV light curve, 
which has a much higher cadence than the IR ones, as  the primary variation, 
and the W1- and W2-band light curves as  the responding variation. 

First, we restricted the $\tau$ range to be 0--1500~days, which is the same as that in the ICCF analysis, 
while no restriction was applied to the width ($w$) range of the top-hat TF.
Top panel in Fig.~\ref{fig:fig4} shows the results of the JAVELIN simulations.
Similar to the results of the ICCF analysis in Fig.~\ref{fig:fig3}, multiple likelihood peaks were found
at  time lags (both $\tau_{\rm W1}$ and $\tau_{\rm W2}$) of $\sim 60$~days, $\sim 130$~days, and $\sim1250$~days. 
Furthermore, we obtained likelihood peaks of $w \sim 0$~days and 300--350~days (inset of the top panel of Fig.~\ref{fig:fig4}).

Then, in order to identify the $\tau$ peaks corresponding to
$w \sim 0$ days and $w \sim 300$--350~days separately,
we performed the JAVELIN simulations where 
the $w$ range was restricted to be 0--100~days and 200--400~days, respectively.
The middle and bottom panels of Fig.~\ref{fig:fig4} show
the resultant respective likelihood distributions. 
We found three peaks in the likelihood distributions of
$\tau_{\rm W1}$ and $\tau_{\rm W2}$ for $w=$0--100~days 
($\tau \sim 60$, $140$, $1280$~days),
 whereas two peaks for $w=$200--400~days
($\tau \sim 130$, $1230$ days).
No significant differences between
$\tau_{\rm W1}$ and $\tau_{\rm W2}$ were found.
 Table~\ref{tab:tab1} lists the resultant $\tau_{\rm W1}$ and $\tau_{\rm W2}$ and their $1\sigma$ errors.
The insets of the middle and bottom panels of Fig.~\ref{fig:fig4} show
the likelihood distributions of the widths of the top-hat TF 
in the W1 band ($w_{\rm W1}$) and W2 band ($w_{\rm W2}$).
 Their values were estimated to be 
 $8.2^{+16.7}_{-6.3}$ and
 $9.4^{+20.4}_{-7.3}$~days for $w_{\rm W1}$ and $w_{\rm W2}$, respectively,
for $w=$0--100 days, and similarly 
 $302.6^{+17.8}_{-26.9}$  and
 $326.9^{+13.8}_{-20.9}$ days, respectively, 
for $w=$200--400 days.

\renewcommand{\arraystretch}{1.1}
\begin{table}
 \caption{Detected $\tau$ of the W1 and W2 bands behind the X-ray data measured  with the JAVELIN simulations. 
 Errors refer to $1\sigma$ uncertainties.}
 \label{all_tbl}
 \begin{center}
  \begin{tabular}{ccc}
   \hline\hline
 $w$ range~(days)&$\tau_{\rm W1}$~(days)  & $\tau_{\rm W2}$~(days) \\\hline
 0--100  & $63.3^{+8.8}_{-8.1}$ & $63.5^{+9.1}_{-9.2}$\\ 
  &$135.5^{+15.6}_{-13.4}$ & $143.8^{+20.2}_{-14.0}$\\ 
  &$1281.4^{+4.8}_{-6.3}$ & $1281.2^{+5.1}_{-5.6}$\\[1.5ex]
  200--400 & $127.5^{+14.0}_{-13.7}$ & $132.9^{+13.6}_{-11.8}$\\ 
    &$1226.6^{+11.9}_{-9.3}$ & $1228.0^{+10.8}_{-9.2}$\\
\hline\hline
  \end{tabular}
\end{center}
\label{tab:tab1}
\end{table}

\begin{figure*}
	\includegraphics[width=160mm]{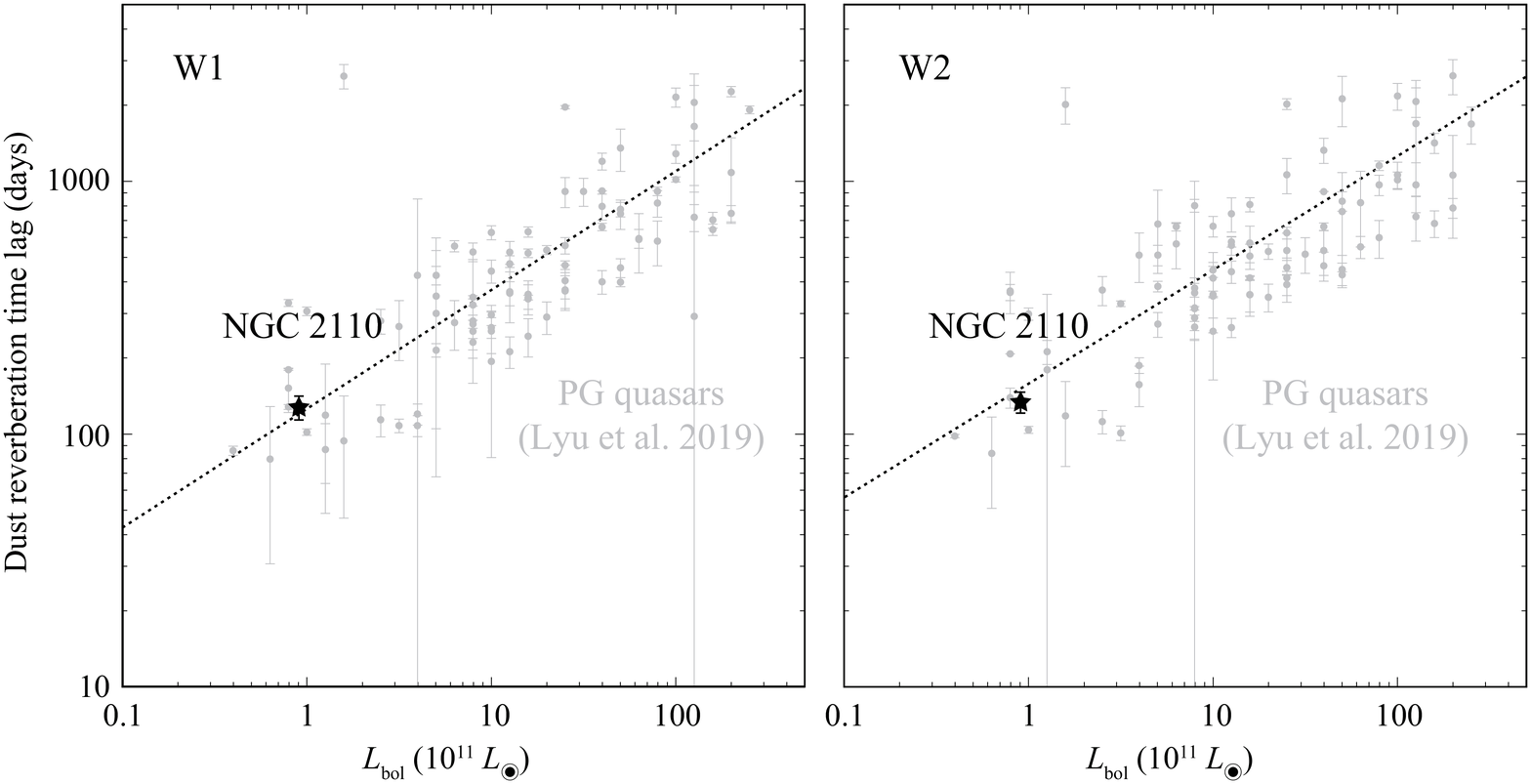}
    \caption{Left panel shows the correlation between $\tau_{\rm W1}$ and $L_{\rm bol}$ of NGC~2110 and PG quasars from Lyu et al. (2019),  whereas right panel shows  that between $\tau_{\rm W2}$ and $L_{\rm bol}$. The black star in each panel shows  the data point for NGC~2110   for  $\tau_{\rm W1} \sim\tau_{\rm W2} \sim130$~days, and  grey circles show those of PG quasars. Error bars do not include the uncertainty of the NGC~2110 luminosity  due to that of the bolometric correction for the X-ray luminosity.}
    \label{fig:fig7}
\end{figure*}

\subsection{Examining Modelled Light Curves}

The JAVELIN simulation calculate the modelled light curve, 
which is the expected average of all possible light curves, 
and its uncertainty, which is the variance of them,
for both the source (X-ray) and response (IR) flux variations.
The next  step is to find which likelihood distribution peak of $(\tau ,\; w)$ 
is the most favoured  from these best-fitting modelled light curves.
For this purpose, we performed additional JAVELIN simulations
for each likelihood peak of $\tau$ restricting the $w$ range within 0--100 days
 and 200--400 days.

Figure~\ref{fig:fig5}(a) and (b) show the modelled light curves
for $(\tau,\; w)\sim (60,\; 12)$ and $(140,\; 12)$~days, respectively. 
At first glance, both IR and X-ray modelled light curves well reproduce 
the observed flux data, including even flux variations on timescales of $\sim$100 days. 
However, a closer look at both the IR modelled light-curves reveals that remarkable flux variations 
 happen  in many of the half-year gaps of  the observed epochs during 
 $\sim 57800$--58500~MJD, while none of the observed points 
(Fig.~\ref{fig:fig1}) show much variation. 
In other words, most of sharp positive peaks conveniently fall between observed epochs.
The sharp dip at around $\sim 57450$~MJD which splits a major peak 
(corresponding to peak A) in the IR modelled light-curve
in Fig. \ref{fig:fig5}(a) upper panels appears rather artificial.
In addition, it has been known that  the AGN central engine usually shows
a smaller flux variation in shorter timescales, and moreover that 
the IR responses to its short-term flux variations  are smeared 
 regardless of the geometry of the IR emitting region unless the viewing angle is perfectly face-on
(e.g., \citealt{1992ApJ...400..502B}; \citealt{2019ApJ...886...33L}). 
This fact suggests short-term variations in the IR light-curve as in the modelled light-curves 
to be even less realistic. 
Therefore, the remarkable IR flux variations  that happens ``conveniently'' 
in gaps of the  observed epochs in the modelled light-curves 
are, though one of the mathematical solutions, unlikely to be the real case.    

Figure~\ref{fig:fig5}(c) shows the modelled light curves for $(\tau,\; w)\sim (1280,\; 12)$~days.
In this case, the IR modelled light curves show remarkable flux variations on timescales 
of less than half a year in not only similar periods of observation gaps 
as in the previous two cases but also during $\sim 56700$--57300~MJD, 
 where again none of the observed points show much variation. 
  We interpreted that the distinctive variation during $\sim 56700$--57300~MJD, 
  which is characterized with two sharp peaks coinciding at the two observed 
  epochs in each IR band, originated  in the mismatch between the IR flux level  during the period 
 and the X-ray flux level 1280~days prior to it, i.e. during $\sim 55400$--56000~MJD.
More specifically, since the errors of the observed IR flux data  are very small,
the IR modelled  fluxes at the observing epochs  are inevitably constrained to be 
very close to  the observed data,  whereas  at any other epochs they  remain faint, 
following the X-ray flux data points.
We  conclude that these strong short-term flux variations  in IR and the sharp peaks in X-rays
in the modelled light curves during these epochs are not realistic.

Figures~\ref{fig:fig6}(a) and (b) show the modelled light curves
for $(\tau,\; w)\sim (130,\; 310)$ and $(1260,\; 340)$~days, respectively.
In both cases, we found that the IR modelled light curves connected the observed 
data points smoothly.
In the former case ($(\tau,\; w)\sim (130,\; 310)$~days), the X-ray modelled light curves 
also successfully follow the observed data points overall for the entire observed period.
In the latter case ($(\tau,\; w)\sim (1260,\; 340)$~days), however,
the X-ray modelled light curves were brighter than the observed data points 
during $\sim 55500$--56000~MJD.
We  interpret that this flux discrepancy  was caused by the same origin as for 
the mismatch between the IR and X-ray flux levels
at the corresponding epochs in the case of $(\tau,\; w) \sim (1280,\; 12)$~days (see previous paragraph).
In this case, the X-ray flux variations in short timescales were smeared out 
by the large-$w$ TF  and accordingly resulted in the smooth IR modelled light curves.
Then, the IR modelled  fluxes  stayed close to the observed data points,
 whereas the X-ray modelled  fluxes did not,
probably because the errors of the IR flux data points were
much smaller than those of the X-ray  ones.

 Finally we conclude that the most favoured parameter set of the dust-reverberation time lags in NGC~2110 are 
$\tau_{\rm W1} = 127.5^{+14.0}_{-13.7}$~days and 
$\tau_{\rm W2} = 132.9^{+13.6}_{-11.8}$~days
with  TF widths of
$w_{\rm W1} \sim 300$~days and 
$w_{\rm W2} \sim 330$~days
(Figure \ref{fig:fig6}a).

\subsection{Comparison of $\tau - L$ relation with type 1 AGNs} 

The dust-reverberation radius is known to be correlated well with the AGN luminosity
for type 1 Seyfert galaxies and quasars, and its origin 
 is well explained by the dust reverberation at  the innermost region of  the dusty torus
(e.g., \citealt{2014ApJ...788..159K}; \citealt{2019ApJ...886..150M}; \citealt{2019ApJ...886...33L}). 
 Here we examine whether or not  the dust reverberation lags and the luminosity of
the type 2 AGN NGC~2110 follow the same trends. 

Since we have obtained the dust-reverberation lags for the {\it WISE} $W1$ and $W2$ 
bands ($\tau_{\rm W1}$ and $\tau_{\rm W2}$),
we compare the  correlation between the dust-reverberation lags in the same bands
and the bolometric luminosity presented by \cite{2019ApJ...886...33L}.
In order to estimate the bolometric luminosity of NGC~2110,
we utilize $L_{\rm 14-195~keV} \sim  4.3 \times 10^{43}$~erg~s$^{-1}$ reported by \cite{2017ApJS..233...17R}, 
and multiply the bolometric correction factor of 8 \citep{2009MNRAS.392.1124V}, following the estimate by \cite{2017ApJ...850...74K}. 
In Fig.~\ref{fig:fig7}, we plot the  dust-reverberation lags ($\tau_{\rm W1}$ and $\tau_{\rm W2}$) and $L_{\rm bol}$ 
of NGC~2110 on those of type-1 PG quasars reported by \cite{2019ApJ...886...33L}. 
The data points of NGC~2110 are found to be located on the best-fitting correlations
for the PG quasars.
 A potential issue that has not been considered is the systematic uncertainty in the bolometric correction,
which is reported to be  approximately $0.5$~dex (\citealt{2009MNRAS.392.1124V}). 
Then we find that even if this uncertainty is incorporated,
the data points of NGC~2110 are still consistent with the correlation for the PG quasars. 
This result indicates that the origin of the NIR flux variation of NGC~2110 is 
the same as in type 1 AGNs; i.e. the NIR emission  responsible for the  flux variation 
comes from the innermost region of the dust torus and  responds to 
the flux variation of the central engine with a lag.

\section{Discussion}

We have measured a dust-reverberation lag in a type 2 Seyfert AGN for the first time;
the time lag between the X-ray and NIR (continuum emission in the W1 and W2 bands) 
flux variations of NGC~2110 was estimated  to be $\tau _{\rm dust} \sim 130$~days.
 This value is consistent with the dust-reverberation radius-luminosity relation
for  type 1 AGNs. This fact suggests that the observed NIR  emission
 originates in the thermal emission of hot dust
in the innermost dust torus.

According to the AGN unified model,  one may expect that the emission 
from the innermost dust torus is totally obscured by the outer dust torus.
However, high angular-resolution integral-field spectroscopies
 reported that some of type 2 AGNs follow the correlation for type 1 AGNs 
between the NIR luminosity and the isotropic luminosity indicators
(such as mid-infrared continuum emission and hard X-ray emission).
 This correlation suggests that a significant amount of the NIR flux
comes from hot dust in the innermost dust torus
in relatively less obscured type-2 AGNs
(\citealt{2015A&A...578A..47B}; \citealt{2018ApJ...858...48M}).
 \citet{2015A&A...578A..47B} reported that NGC~2110 is one of those 
 less obscured type-2 AGNs, supported by its relatively small hydrogen column density
of $N_{\rm H}\sim 5\times 10^{22}$ cm$^{-2}$ (\citealt{2015MNRAS.447..160M}; \citealt{2018ApJ...854...42B}).
 In fact even a Compton-thick (i.e. heavily-obscured) type-2 AGN, NGC~1068,
was reported to show a large amplitude of NIR flux variations 
(\citealt{2004MNRAS.350.1049G}; \citealt{2006A&AT...25..233T})
\footnote{
\cite{2001ApJ...556..694G} explained the loss of the maser signal
and the peak of the NIR light curve of NGC 1068
 with  reverberation of a suppositional flare of the nucleus emission.
}.
 Furthermore, a recent NIR interferometer observation of it found a ring-like structure of emission,
which was associated with the dust sublimation region (\citealt{2020A&A...634A...1G}). 
As such, observations of NGC 1068 revealed that even heavily-obscured type-2 AGNs, 
if not all of them, show NIR nucleus emission, which implies that NIR nucleus emission 
originating probably in the innermost dust torus may be common among any type 2 AGNs, 
let alone less obscured ones like NGC~2110. 
Therefore, it is likely that the observed NIR continuum emission and its flux variation of NGC~2110
originate in the innermost dust torus.

We have estimated the full widths of the TF  to be approximately 
$w_{\rm W1} \sim 300$~days and $w_{\rm W2} \sim 330$~days for the W1 and W2 bands, respectively.  
Thus the TF is considerably extended.
The width of the TF of the torus emission is generally larger when the viewing angle is larger
(e.g., \citealt{1992ApJ...400..502B}; \citealt{2011ApJ...737..105K};
\citealt{2020ApJ...891...26A}). 
In general, type 2 AGNs are considered to have a large viewing angle according to the unified model; 
accordingly the TF of type 2 AGNs should be extended. 
Therefore, our estimated large width of the TF for NGC~2110  is
consistent with the picture of the unified model.

We have used JAVELIN for the lag estimation,
which assumes a top-hat function in the transfer equation,
 and have estimated  the most likely time lag of $\tau _{\rm dust} \sim 130$ days
and  full width of $w \sim 300$--$330$~days for the TF. 
This TF has a positive value of the response lag from $\sim -30$~days to $\sim 290$~days;
 this means that the earliest NIR flux variation starts 
$\sim 30$~days ($\sim 10$\% of the full width of the TF)
before  its triggering X-ray flux variation.
This negative lag response apparently violates causality 
and  therefore should not be real.
We  conjecture that it is caused by
JAVELIN's assumption of the top-hat shape for the TF 
without any constraints between the lag and width in the calculation
and also by the sparse monitoring sampling of the IR flux variations.
Data from monitoring observations with a high cadence combined 
with methods of non-parametric reconstruction of the TF
(e.g., \citealt{1994ASPC...69...23H}; \citealt{1994MNRAS.271..183P}; \citealt{1995ApJ...440..166K})
will enable us to constrain  the shape of the TF
and provide  more detailed information on
the geometry and physical parameters of the innermost dust torus.

The dust reverberation mapping can be applied even to obscured AGNs, 
as we have demonstrated in this work. 
The viewing-angle dependence of the torus TF can be studied by 
a systematic dust reverberation survey on both type 1 and type 2 AGNs, 
which will resolve part of parameter degeneracy in the torus TF. 
   Understanding the dust tori of AGNs  is also  important for investigating
the radius-luminosity relation of AGNs, which is used as a high-redshift distance indicator
(e.g., \citealt{2014ApJ...784L..11Y}; \citealt{2014ApJ...784L...4H}; \citealt{2017ApJ...842L..13K}),
 because the relation might be affected  by the dust torus parameters  in each AGN
(e.g., \citealt{2014ApJ...784L..11Y}; \citealt{2019ApJ...886..150M}).
More systematic studies of type 2 AGNs are  desired
with simultaneous X-ray and NIR  monitoring.
 For example, not only \textit{MAXI} but also \textit{Swift}/BAT
data of type 2 AGNs are useful to be compared  with
NIR data including the \textit{WISE} archive data. 
 Our results will be reported in the forthcoming paper.

Another promising tool to investigate the torus structure is 
X-ray microcalorimeter spectroscopy  of a narrow Fe-K$\alpha $ line,
which will provide  useful information on kinematics in the dust torus 
from its line width and profile, as demonstrated by the study of NGC~1275,
using the Soft X-ray Spectrometer (SXS) on board the Hitomi satellite
(\citealt{2018PASJ...70...13H}).
The next X-ray microcalorimeter mission \textit{XRISM} will be launched 
from Japan Aerospace Exploration Agency (JAXA) in 2021 Japanese fiscal year,
and \textit{Athena} is planned to be launched from European Space Agency (ESA) in 2031.
 With combination of the future X-ray microcalorimeter studies
and the dust reverberation mappings on various types of AGNs,
 as well as IR interferometer observations,
 our understanding of the  innermost torus will be dramatically improved.

\vspace{0.5cm}

\section{Conclusions}

We performed  dust-reverberation mapping on  the type 2 AGN NGC~2110 for the first time, 
 using the 2--20~keV \textit{MAXI} data and the W1- and W2-band  \textit{WISE} data.  
We conducted the ICCF analyses and  JAVELIN simulations to compare them 
and obtained  possible dust-reverberation time lags $\tau_{\rm dust}$ 
for both the W1 and W2 bands   of $\sim 60$~days, $\sim 130$~days and $\sim 1250$~days. 
We moreover examined the X-ray and IR best-fitting modelled light curves obtained by the JAVELIN runs. 
As a result, we revealed that the time lag of  
 $\tau_{\rm dust} \sim 130$~days is most favoured. 
By employing the time lag of  $\tau_{\rm dust} \sim 130$~days,
the  $\tau_{\rm dust}$--$L_{\rm bol}$ relation of NGC~2110 was found to be consistent with 
those in type 1 AGNs. 
This means that the dust reverberation originates in hot dust around the innermost region of the dusty torus 
in the type-2 NGC~2110 as in the type 1 AGN. 
The present study demonstrated that X-ray and IR simultaneous monitoring  is a powerful  method 
to conduct dust-reverberation mapping on type 2 AGNs. 
To combine them with observations with  future X-ray microcalorimeter missions will enable 
us to  gain deeper insight about  the innermost regions of AGN tori.

\section*{Acknowledgements}
We thank the anonymous referee for his/her valuable suggestions and comments. 
We are grateful  for prof. Martin Ward for his helpful comments. 
We thank Dr. Nakahira for his useful comments about the \textit{MAXI} data.
This research has made use of \textit{MAXI} data provided by RIKEN, JAXA and the \textit{MAXI} team.
This publication makes use of data products from the Wide-field Infrared Survey Explorer, 
which is a joint project of the University of California, Los Angeles, 
and the Jet Propulsion Laboratory/California Institute of Technology, 
funded by the National Aeronautics and Space Administration.
This research is supported by Japan Society for the Promotion of Science (JSPS) KAKENHI with the Grant number 
of 16H02162, 17J09016, 19K21884, and 20H01947. 
Part of this research was financially supported  with the Grant-in-Aid for JSPS Fellows 
for young researchers (T.K. and M.K.).







%
%
%
%

\bsp	
\label{lastpage}
\end{document}